%% file: paper.tex
\RequirePackage{fix-cm}
\documentclass[twocolumn,epjc3]{svjour3}  
\smartqed  
\RequirePackage{graphicx}
\RequirePackage{physics}
\RequirePackage{cite}
\RequirePackage{slashed}
\RequirePackage{subcaption}
\RequirePackage{fontawesome5}
\RequirePackage[table]{xcolor}
\RequirePackage{tikz-feynman}
\RequirePackage{mathtools}
\RequirePackage[utf8]{inputenc}
\RequirePackage[T1]{fontenc}
\RequirePackage{listings}
\RequirePackage{textcomp}

\definecolor{codegreen}{rgb}{0,0.6,0}
\definecolor{codegray}{rgb}{0.5,0.5,0.5}
\definecolor{codepurple}{rgb}{0.58,0,0.82}
\definecolor{backcolor}{rgb}{0.95,0.95,0.92}
\definecolor{tabgrey}{rgb}{0.57,0.57,0.57}

\lstdefinestyle{mystyle}{
    backgroundcolor=\color{backcolor},   
    commentstyle=\color{codegreen},
    keywordstyle=\color{magenta},
    numberstyle=\tiny\color{codegray},
    stringstyle=\color{codepurple},
    basicstyle=\ttfamily\footnotesize,
    breakatwhitespace=false,         
    breaklines=true,                 
    captionpos=b,                    
    keepspaces=true,                 
    numbers=left,                    
    numbersep=5pt,                  
    showspaces=false,                
    showstringspaces=false,
    showtabs=false,                  
    tabsize=4
}
\newcommand{\madnlo}{{\sc MadGraph5\_aMC@NLO}}
\newcommand{\mg}{MG5aMC}
\newcommand{\mgnlo}{MG5aMC}
\newcommand{\aloha}{{\sc Aloha}}
\journalname{Eur. Phys. J. C}
\begin{document}

\title{Speeding up MadGraph5\_aMC@NLO

}


\author{O. Mattelaer\thanksref{e1,addr1,addr2}
        \and
        K. Ostrolenk,\thanksref{e2,addr1,addr3} 
}

\thankstext{e1}{e-mail: olivier.mattelaer@uclouvain.be}
\thankstext{e2}{e-mail: kiran.ostrolenk@manchester.ac.uk}

\institute{ Center for Cosmology, Particle Physics and Phenomenology (CP3), Universit\'e Catholique de Louvain (UCLouvain), B1348 Louvain-la-Neuve, Belgium \label{addr1}
           \and
           Centre de Calcul Intensif et de Stockage de Masse, Universit\'e Catholique de Louvain (UCLouvain), B1348 Louvain-la-Neuve, Belgium \label{addr2}
           \and
          University of Manchester, School of Physics and Astronomy, Schuster Building, Oxford Road, Manchester M13 9PL, United Kingdom \label{addr3}
}

\date{Received: date / Accepted: date}

\maketitle

\vspace*{-7.9cm}
\noindent
\small{MCNET-21-01, CP3-21-01, MAN/HEP/2021/001}
\vspace*{8.1cm}
\begin{abstract}
In this paper we will describe two new optimisations implemented in \madnlo, both of which are designed to speed-up the computation of leading-order processes (for any model).
First we implement a new method to evaluate the squared matrix element, dubbed helicity recycling, which results in factor of two speed-up. 
Second, we have modified the multi-channel handling of the phase-space integrator providing tremendous speed-up for VBF-like processes (up to thousands times faster).

\end{abstract}

\section{Introduction}
\label{intro}

While the LHC is still running, preparation is starting for the High-Luminosity LHC. As part of this preparation, the CPU efficiency of our Monte-Carlo generators is crucial due to the sheer number of events that need to be generated. Given the current constraints on the LHC IT budget this will not be possible without significant software improvement \cite{Aarrestad:2020ngo,Alves:2017she}. While the full CPU time of the LHC experiment is not dominated by event generation, it is still estimated to represent between ten and twenty percent of it. Consequently, we have received a specific request to speed-up that step by at least  20\% and ideally by a factor of two \cite{Aarrestad:2020ngo}.

In addition to the High-Luminosity LHC, work is also starting for future high-energy accelerators \cite{EuropeanStrategyGroup:2020pow}. Going to the high-energy regime will amplify multi-scale issues which in turn can reduce the efficiency of event generation. This issue is particularly striking within  \madnlo\ \cite{Alwall:2014hca,Alwall:2011uj} (\mgnlo) for VBF-like processes where the current phase-space integration algorithm either fails to generate the requested number of events or takes an unreasonable time to do so.

The different avenues for speeding up Monte-Carlo integration is a well covered topic. 
 Such efforts can be classified into four different categories. First one can optimise the evaluation of the function being integrated, which in our case is the matrix element \cite{Alnefjord:2020xqr,Maltoni:2002mq,Berends:1987me,Berends:1988yn}. Second one can optimise the integration method to minimise the number of times such functions need to be evaluated \cite{Gao:2020zvv,Bendavid:2017zhk,Bothmann:2020ywa,Klimek:2018mza,Backes:2020vka}. Third, one can try to use more efficiently the various types of hardware (e.g. GPU, MPI, vectorization) \cite{Hagiwara:2013oka,Hagiwara:2009cy,Hagiwara:2009aq,Benjamin:2017xdd} and finally one can play with the weights of the sample to optimise/re-use information \cite{Matchev:2020jqz,Nachman:2020fff,Frederix:2020trv,Mattelaer:2016gcx,Andersen:2020sjs, Brooks:2020bhi}.
 
In the context of this work, we will focus on optimising \mgnlo, one of the main Monte-Carlo generators \cite{Frixione:2007vw,Bothmann:2019yzt,Kilian:2007gr} and we will combine two different methods to achieve our goal, one optimising the time to evaluate the matrix element and one optimising the phase-space integrator. The strategy we employed was to keep the main design choices in place (in particular the helicity amplitude method \cite{Murayama:1992gi} and the single diagram enhancement method\cite{Maltoni:2002qb}) and to deeply study them to see how they could be further improved. 

To reduce the time needed to evaluate a given squared matrix element, we use a quite standard Memory/CPU trade-off guided by physics considerations. We have identified parts of the computation that can be stored in memory in order to avoid their re-computation later. We dubbed this method helicity recycling since the additional terms stored in memory correspond to identical sub-expressions shared between different helicity configurations.
 This improvement will be presented in section (\ref{sec:hel}). We start in section (\ref{sec:hel_mg}) by presenting the helicity formalism used in \mgnlo, then we continue in section (\ref{sec: hel recyc}) by explaining the main idea behind helicity recycling. The details of the associated speed-up will be then presented in section (\ref{sec: hel result}). 

The second improvement that we have implemented improves the phase-space integrator. Contrary to the first method, it is more challenging to introduce an improvement that acts positively on all processes. On the other hand, the expected (and obtained) gain can be much more impressive with this method. For this paper we will mainly focus on the speed-up of VBF-like processes since they are the ones where \mgnlo\ has some specific issues. 

This will be covered in section (\ref{sec:ps}) where we start  by reviewing the current multi-channel strategy (the single diagram enhancement method) in section (\ref{sec:sdereview}). 
We then explain in section (\ref{sec:tchannel}) the methods used to have a better handling on the integration of $t$-channel propagators.
 We then turn in section (\ref{sec:propa}) to the deeper changes we made to the multi-channel strategy as well as the impact on some physical observables -- which are beyond LO accurate. Our speed-ups are then compared to the older version of the code in section (\ref{sec:psresult}).

Our conclusions will be presented in section (\ref{sec:conclusion}). We also provide two appendices. First,  we will give a  brief manual on how to tune the optimisation parameters. Finally in \ref{sec:aloha}, we describe the modifications (and the associated conventions)  to the \aloha\ package\cite{deAquino:2011ub}, related to the helicity recycling algorithm.

\section{Helicity recycling within the Helicity Amplitude Method}
\label{sec:hel}
\subsection{Helicity Amplitudes and previous optimisations in \mg}
\label{sec:hel_mg}

When evaluating a matrix element one can identify two key structures: the Lorentz structure of the matrix element and its colour structure. Within \mg, the evaluation of these two structures factorises at the amplitude level  \cite{Maltoni:2002mq}. Hence it is possible to discuss one without the other. The helicity-recycling optimisation only impacts the evaluation of the Lorentz structure and in this section we will explain how this evaluation is performed. 

\subsubsection{Helicity amplitude formalism}
\label{sec: HAF}

Matrix elements typically contain factors of spinors (from the external fermions) and Lorentz vectors (from internal propagators). When analytically evaluating the square of this matrix element it is common to remove the dependence on spinors  via their on-shell condition:

\begin{equation}
    \sum_{s=\pm1} u_s(p)\bar{u}_s(p)=\slashed{p} + m,
\end{equation}

\noindent where the sum is over helicity. The squared matrix element can then be reduced to a series of scalar products between Lorentz vectors. 
However, such a formalism is typically not used by programs that perform numerical evaluations of matrix elements. This is because such a formalism will cause the computational complexity to grow quadratically with the number of Feynman diagrams (due to the presence of interference terms).\footnote{This strategy is actually a winning one at  low-multiplicity and is used for example in {\tt CalcHEP}~\cite{Belyaev:2012qa}.}

One solution is to use the helicity amplitude formalism \cite{DeCausmaecker:1981wzb,DeCausmaecker:1981jtq,Gastmans:1989wb} where the summation over helicity is postponed. Under this formalism the matrix element is reduced to a series of spinor products, rather than Lorentz products.\footnote{This is in general possible because $\text{SU(2)}\times\text{SU(2)}$ is the double cover of SO(1,3).} These spinors will depend on the specific helicities of the external particles. The advantage of this formalism is that its complexity grows linearly with the number of diagrams (since interference terms can be handled by a simple sum over amplitudes).\footnote{The recursion relation method \cite{Britto:2004ap,Britto:2005fq} allows for an even faster evaluation of matrix elements but such a method is not compatible with the phase-space integrator of \mg.}

\begin{figure}
  \centering
  \includegraphics{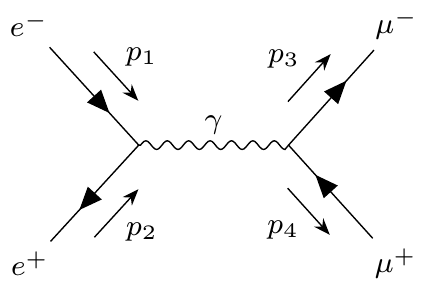}
  \caption{$S$-channel \(e^-e^+\rightarrow \mu^- \mu^+\) diagram with photon mediator. We use $p_1$, $p_2$, $p_3$ and $p_4$ to label the external momenta.}
  \label{fig: ee2mumu feyn}
\end{figure}

An example will help illustrate the key principles here. Following Ref. \cite{Schwartz:2013pla}, let us work in the massless limit and consider the unpolarised $e^-e^+\rightarrow \mu^- \mu^+$ $s$-channel matrix element with a photon mediator. Let us use $p_1,p_2$ to label the momenta of the electrons and $p_3,p_4$ for the muons. This process is pictured in Fig. (\ref{fig: ee2mumu feyn}).\footnote{The Feynman diagrams in this paper were generated using the Tikz-Feynman package \cite{Ellis_2017}.} 

In what follows it will also be helpful to use the standard Mandelstam variables, defined as:
\begin{align}
    s &= (p_1+p_2)^2 = (p_3+p_4)^2\\
    t &= (p_1-p_3)^2 = (p_4-p_2)^2\\
    u &= (p_1-p_4)^2 = (p_3-p_2)^2
\end{align}

\noindent Accordingly, this matrix element is written as:

\begin{equation}
    i\mathcal{M} = (-ie)^2\bar{v}(p_2)\gamma_\mu u(p_1) \frac{-ig^{\mu \nu}}{s} \bar{u}(p_3) \gamma_\nu v(p_4).
\end{equation}

\noindent Remember that $u(p)$ and $v(p)$ are Dirac spinors and can either be left- or right-handed. In the massless limit this corresponds to them having $+1$ or $-1$ helicity eigenvalues respectively. In what follows we will also use $\psi$ to denote a general Dirac spinor. The subscript $L$ and $R$ will be used to denote left- and right-handedness respectively. It is common to use the following notation when expressing helicity amplitudes, for a spinor of momentum $p_n$:

\begin{align}
    \psi_L = n\rangle &, \quad \psi_R = n], \\
    \bar{\psi}_L = \langle n &, \quad \bar{\psi}_R = [ n.
\end{align}

One can show that \([n \gamma^\mu m] = \langle n \gamma^\mu m\rangle = 0\). Hence, for the matrix element to be non-zero, the incoming fermions must have opposite helicity and similarly for the outgoing. Hence, for all the possible helicity combinations we could construct, the only ones that give a non-zero matrix element are:

\begin{align}
\label{eq: hc 1}  &1^-2^+3^-4^+, \\
\label{eq: hc 2}  &1^+2^-3^+4^-, \\
\label{eq: hc 3}  &1^+2^-3^-4^+, \\
\label{eq: hc 4}  &1^-2^+3^+4^-.
\end{align}

\noindent Here, the notation $n^m$ means the fermion with momentum $p_n$ has helicity $m$. The first helicity combination will give:

\begin{align}\label{eq: hel amp}
    i\mathcal{M}(1^-2^+3^-4^+)&=(-ie)^2\langle 2 \gamma_\mu 1] \frac{-ig^{\mu \nu}}{s}\langle 3 \gamma_\nu 4], \\
                              &=2\frac{ie^2}{s}[41]\langle 23 \rangle. \label{eq: hel amplit}
\end{align}

%

Using the relation $[nm]\langle mn \rangle = \langle m n \rangle [nm] = p_m \cdot p_n $, one finds the square to be:

\begin{equation}
    \abs{\mathcal{M}(1^-2^+3^-4^+)}^2=4e^4\frac{u^2}{s^2}, \label{eq: sqrd hel amp}
\end{equation}

\noindent By parity, the helicity combination $1^+2^-3^+4^-$ will give the same result. The only remaining contributing helicity combinations are the same two but with $1 \leftrightarrow 2 $. By symmetry arguments, one can see they will give the same result as in equation (\ref{eq: sqrd hel amp}) but with $u \rightarrow t$. Hence the final result for this matrix element becomes:

\begin{equation}
    \frac{1}{4}\sum_{\text{helicity}}\abs{\mathcal{M}}^2=2e^4\frac{t^2+u^2}{s^2}.
\end{equation}


\subsubsection{{\sc MadGraph} implementation} 
\label{sec:MadGraph_impl}
The main points from the simple example looked at in section (\ref{sec: HAF}) are:

\begin{enumerate}
    \item Diagrams are evaluated at the amplitude level (before squaring the matrix element).
    \item To get the final result these diagrams must be summed over helicity combinations.
    \item Only some of these helicity combinations will actually contribute.
\end{enumerate}

\begin{figure*}
  \input{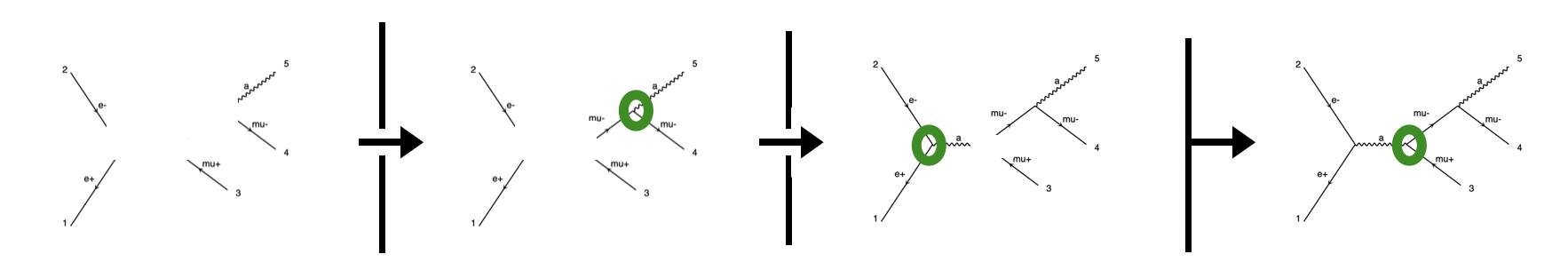}
  \caption{Iterative steps used to evaluate a given Feynman diagram in the helicity amplitude formalism. In (a) the spinors of the external particles are evaluated. In (b) and (c) the algorithm works through each vertex (circled), evaluating the wave function of the associated propagator (in red). Finally in (d) the last vertex is reach at which point the algorithm uses what it has calculated so far to evaluate the amplitude.}
  \label{fig:hel_formalism}
\end{figure*}

\mg\ follows the HELAS strategy \cite{Murayama:1992gi} to compute amplitudes. It starts by computing the spinorial representation of all external particles. These spinors are then iteratively combined using some model specific functions (generated thanks to the \aloha \cite{deAquino:2011ub} package). In this way the full diagram can be built up. This process is represented in Fig. (\ref{fig:hel_formalism}). In the helicity amplitude formalism one distinguishes three types of functions:
\begin{itemize}
    \item {\bf external wave function:} function calculating the spinorial representation (e.g. $u$, $\bar v$, $\epsilon^\mu$) of the external particles evaluated for a given helicity.
    \item {\bf internal wave function:} function calculating the spinorial representation (see Eq. (\ref{eq:photon_wfct})) of an internal particle (i.e. of a propagator) evaluated for a given helicity combination.
    \item {\bf amplitude:} function fully contracting all spinorial representations and therefore returning the value for a given helicity combination of the amplitude of the associated Feynman diagram.
\end{itemize}

\noindent As depicted in Fig. (\ref{fig:hel_formalism}), the evaluation of the matrix element starts by evaluating all the external wave functions, before combining them to get the internal wave functions associated to the propagators.  Finally it reaches the last vertex and at that point returns an amplitude.


For example, consider again the $e^-e^+\rightarrow \mu^- \mu^+$ process.
After the computation of the spinors associated to the electrons and muons (which depends only on their momenta and helicity), the algorithm will call the routine for calculating the internal photon wave function. The analytic expression of this photon wave function $\phi_\gamma$ will be:
 \begin{equation}
    \phi_\gamma^{\mu h_3h_4} = -ie\frac{-ig^{\mu \nu}}{s}\bar{\psi}^{h_3}_{\mu^+} \gamma_\nu \psi_{\mu^-}^{h_4},   
    \label{eq:photon_wfct}
 \end{equation}

\noindent where $\psi_\mu$ has been used to represent the muon spinors. As already mentioned, these will be dependent on the helicity of the muons, labelled $h_3$ and $h_4$.
Note that the wave function associated to this propagator is not unique since it depends on which vertex is used to compute it.


In this example, we are already at the last vertex and all that remains is to contract the various wave functions, taking into account the Lorentz/spinor structure of the associated vertex. Analytically, this is written as:
\begin{align}
    \mathcal{M}^{h_1h_2h_3h_4} &= \left(-ie\bar \psi_{e^-}^{h_2} \gamma_\mu \psi_{e^+}^{h_1} \right)\phi_\gamma^{\mu h_3h_4}, \label{eq: this is a ref} \\
     &= (-ie)^2\bar{\psi}_{e^-}^{h_2} \gamma_\mu \psi_{e^+}^{h_1} \left(\frac{-ig^{\mu \nu}}{s}\bar{\psi}_{\mu^+}^{h_3} \gamma_\nu \psi_{\mu^-}^{h_4}\right),
\end{align}

\noindent where, just as with the muons, $\psi_e$ has been used to represent the spinors of the electrons. This is the same as the expression in (\ref{eq: hel amp}) but without the specific choice of helicity combination.

\mg\ generates a FORTRAN subroutine that carries out the above process for a given matrix element and helicity combination and then returns the squared amplitude.\footnote{Evaluation of the matrix element for a single phase-space point is also possible in C++, CUDA and a python wrapper is also available.} \mg\ then performs a loop over all possible helicity combinations, at each point calling this matrix-element subroutine. The results from this loop are then summed to produce the final matrix element. This can be represented, for our $e^-e^+\rightarrow \mu^- \mu^+$ example, by the following pseudo-code:

\begin{lstlisting}[language=Bash]
result = 0
loop over helicity:
|  # external wave functions
|  ext_1 = get_spinor(p_1, hel_1) # e-
|  ext_2 = get_spinor(p_2, hel_2) # e+
|  ext_3 = get_spinor(p_3, hel_3) # mu-
|  ext_4 = get_spinor(p_4, hel_4) # mu+
|      
|  # propagators / internal wave functions    
|  photon = Gamma_mu(ext_3, ext_4) # Eq.(16)
|    
|  # amplitude/final vertex
|  M_diag = Gamma_mu(ext_1, ext_2, photon)
|  result += M_diag^* M_diag
\end{lstlisting}

\noindent which can be generalised for any process as:\footnote{This representation/meta-code is technically accurate only for colourless processes but the general idea is still valid for QCD processes since \mgnlo\ is using the colour-flow formalism \cite{Maltoni:2002mq}.}
\begin{lstlisting}[language=Bash]
result = 0
loop over helicity
|  sumamp = 0
|  loop over all Feynam Diagram
|  |  loop over external particles
|  |  |  ext_i = get_spinor(p_i, hel_i)
|  |          
|  |  loop over propagators
|  |  |  propa_i = Vertex(ext_x, propa_y)
|  |   
|  |  # amplitude/ final vertex  
|  |  sumamp += Vertex(ext_x, propa_y)
|  result += sumamp^* sumamp
\end{lstlisting}

Obviously, \mg\ has already implemented a couple of optimisations. First,  none of the external wave functions depend on the Feynman diagram representation and can therefore be moved outside of the loop over diagrams. Furthermore, the same propagator can appear in multiple Feynman diagrams and in such a case it is also highly beneficial to move it out of the Feynman diagram's loop. Therefore a more accurate representation of \mgnlo, prior to this paper, can be written like:
\begin{lstlisting}[language=Bash]
result = 0
loop over CONTRIBUTING helicity
|  loop over external particles
|  |   ext_i = get_spinor(p_i, hel_i)
| 
|  loop over all possible propagator
|  |   propa_i = Vertex(ext_x, propa_y)
|  
|  # amplitude/ final vertex 
|  sumamp = 0
|  loop over all Feynam Diagram
|  |   sumamp += Vertex(ext_x, propa_y)
|  
|  # squaring the sum of amplitudes
|  result += sumamp^* sumamp
\end{lstlisting}

Secondly, recall the point made at the start of this section that only a subset of helicity combinations contributes. Hence, another optimisation already included in the above meta-code exploits this fact and makes sure the loop over helicity combinations only includes ``contributing'' ones. We shall refer to this optimisation as `helicity filtering'.\footnote{This filtering is done numerically after the numerical evaluations of a couple of phase-space points.}


The matrix routine discussed in this section is called by the \texttt{madevent} binary. This binary is also responsible for generating phase-space points, evaluating the PDFs and writing the events (amongst other things). We can evaluate the computational cost of the matrix routine by comparing the number of instructions it executes to the total number executed by \texttt{madevent}. This is presented in table (\ref{tab: instr nohel}) for top-quark pair production plus gluons.
The diagrams for $t\bar{t}$ are simpler and fewer in number than those of $t\bar{t}gg$ and even more so than those of $t\bar{t}ggg$. Hence the total number of \texttt{madevent} instructions executed by \texttt{madevent} increases across these processes. Furthermore, this means the matrix routine is also responsible for an increasing percentage of the instructions: $23\% \to 96\% \to {\sim}100\%$. 

We also see that for $t\bar{t}gg$ and $t\bar{t}ggg$ (the more complex processes) the amplitude routines account for 44\% and 51\% of the computation, making it the dominant contribution to the matrix element calculation. This is again due to the higher number of diagrams and since the number of unique propagators does not scale as fast as the diagram multiplicity. Hence it is important that any future optimisation targets not just wave-function routines but also the amplitude routines.

Finally, one can also see from the table that the wave-function and amplitude routines do not add up to the number of matrix routine instructions. This is because the routine has other things to evaluate, most noticeably the colour factors for the relevant amplitudes. Such computation is even the second hot-spot for the three gluon multiplicity case and therefore will limit the potential impact of our optimisation.

\input{input/instr_nohel}
\subsection{Helicity recycling}
\label{sec: hel recyc}
In general, when summing over helicity combinations, the same spinor with the same helicity can appear multiple times. For example, in the combinations (\ref{eq: hc 1}) - (\ref{eq: hc 4}) each helicity spinor (such as $1^+$) appears twice. Hence when \mg\ loops over the combinations it will calculate the same wave function for these spinors multiple times (see the above meta-code). This is a waste of computation time (even if in this simple case the external wave functions are cheap to evaluate). It would be more efficient to only calculate the wave functions once, save their values and reuse it when appropriate.

The same principle also applies to the wave function of internal particles. Such routines take other wave functions as input, therefore for a subset of helicity combinations the same input will be given and the code will waste time by re-computing the same quantity.
For example, when looking at the internal (photon) wave function given in Eq. (\ref{eq:photon_wfct}) the helicity combination (\ref{eq: hc 1}) and (\ref{eq: hc 3}) will give exactly the same result. Hence, the computational efficiency can again be improved by only calculating such internal wave functions once and then re-using the value when necessary.

This technique of calling the wave-function routines only once is what we refer to as ``helicity recycling'' and can be written in terms of pseudo-code as: 
\begin{lstlisting}[language=Bash]
loop over external particles
| loop over helicity
| | ext_ij = get_spinor(p_i, hel_j)
 
loop over all possible propagator 
|  loop over helicity combination
|  |   propa_kl = Vertex(ext_ij, propa_xy)

result = 0 
loop over all helicity combination
|  sumamp = 0
|  loop over all Feynam Diagram
|  | sumamp += Vertex(ext_xy,...)
| # squaring the sum of amplitudes
| result += sumamp^* sumamp
\end{lstlisting}

\noindent Here you can note that all stored variables have an additional index indicating which helicity combination was used to compute them.

While such an optimisation sounds natural, it has two potential issues.
First, the amount of RAM needed for the computation will increase. However the RAM is currently dominated by meta-data related to the phase-space integrator and not to the amount of RAM used in the matrix element. The increase of memory needed by the helicity recycling is actually shadowed by such meta-data and hence we did not observe a sizeable increase and certainly not faced any issues with RAM assignment even for the most complex processes.

Second, while the previous strategy was allowing helicity filtering at run time, this method requires us to know which helicity combinations do not contribute when creating the code.
In order to numerically determine the null-helicities, we have designed the code's work-flow in the following way:
\begin{enumerate}
    \item We first allow \mg\ to create the matrix-element subroutine as it normally would.
    \item We then sample a couple of events in order to determine which helicity combinations and amplitudes do not contribute.
    \item Next, the matrix-element subroutine is rewritten in the new paradigm.
\end{enumerate}

The conversion of the code is done by using a directed acyclic graph to represent how the various wave-function and amplitude routines depend on one another. Having established this graph, the program is then able to easily evaluate which wave-function routines are needed to calculate a given diagram. This allows us to avoid duplicate calls to these routines when unrolling the helicity loop. It also allows us to prune efficiently any wave-function (and amplitude) calls that are only associated with vanishing amplitudes.\footnote{These vanishing terms are determined during the initial sampling of the matrix element.}. Compared to the helicity filtering discussed in section (\ref{sec:MadGraph_impl}), this new algorithm is more efficient since it can identify \textit{any} non-contributing component of the computation, like an amplitude that is vanishing only for one particular helicity combination. 

So far we have discussed how helicity recycling is used to minimise calls to external and internal wave functions. However the impact of such an optimisation is at best quite limited since the computation is actually dominated by the amplitude routines.
Thankfully, this implementation also allows us to optimise these routines by a significant factor. 

For the sake of example (and without any lack of generality) let us assume that the final vertex of a given Feynman diagram is a Standard Model fermion-fermion-vector vertex. This corresponds to (see Eq. (\ref{eq: this is a ref})):
\begin{equation}
    \mathcal{M}_{h_1h_2h_\phi} = \bar{\psi}^{h_1}_1\, \gamma_\mu\, \psi^{h_2}_2 \phi^\mu_{h_\phi}.
\end{equation}
Where we have explicitly added indices representing the associated helicity (or helicity combination for internal wave functions) of each component.
Now, when computing $\mathcal{M}_{h_1h_2h_\phi}$ and $\mathcal{M}_{h_1h_2\tilde{h}_\phi}$, the factor $\bar{\psi}^{h_1}_1\, \gamma_\mu\, \psi^{h_2}_2$ will be identical and can therefore be re-used multiple times. Therefore we have been able to optimise the code further by implementing a recycling of this factor. However, a similar factor can also be recycled between $\mathcal{M}_{h_1h_2h_\phi}$ and $\mathcal{M}_{\tilde{h}_1h_2h_\phi}$. Therefore it is advantageous to compare 
(for each Feynman diagram) which one of the following expressions:
\begin{eqnarray}
    &&\bar{\psi}^{h_1}_1\, \gamma_\mu\, \psi^{h_2}_2, \label{aloha_1}\\
    &&\gamma_\mu\, \psi^{h_2}_2 \phi^\mu_{h_\phi} ,\label{aloha_2}\\
    &&\bar{\psi}^{h_1}_1\, \gamma_\mu\, \phi^\mu_{h_\phi}\label{aloha_3},
\end{eqnarray}
can be re-used at a higher frequency and use the most optimal recycling strategy. This optimisation requires us to define a new type of helicity routine and the details of this implementation into \aloha\ are presented in \ref{sec:aloha}.

\subsection{Result}
\label{sec: hel result}

In this section we will quantify the speed-up resulting from using the improvements detailed in section (\ref{sec: hel recyc}). 

\subsubsection{Matrix routine breakdown} 
\label{sub:me break}

\input{input/instr_hel}

First we reproduce table (\ref{tab: instr nohel}) with helicity recycling switched on. This is shown in table (\ref{tab: instr hel}). One can see that for all processes the total number of evaluated instructions has reduced:

\begin{itemize}
  \item from 13G to 11G for $t\bar{t}$ (15\% reduction)
  \item from 470 to 180G for $t\bar{t}gg$ (62\% reduction)
  \item from 11T to 5T for $t\bar{t}ggg$ (55\% reduction)
\end{itemize}

\noindent The latter reductions are much larger because evaluating the matrix element represents a larger percentage of the overall computation for those processes. This is because the diagrams are more complex and numerous for $t\bar{t}gg$ and $t\bar{t}ggg$.

Looking at table (\ref{tab: instr hel}), we observe that both external and internal wave-function routines represent, after helicity recycling,  a relatively insignificant computational cost. Firstly, they were not that significant before the optimisation and secondly they have been highly reduced by the helicity recycling (by at least $10\times$ factor). The final speed-up is actually more dependent on the reduction in calls to amplitude routines. Focusing on the $t\bar{t}gg$ process, one can see the amplitude routines have seen a large reduction in the number of associated instructions (by a factor of two) but still represent 42\% of the overall computation. Although not shown in this table, roughly half of this computation (19\% of the total) is spent evaluating simply scalar products (the contraction of Eq. (\ref{aloha_1}-\ref{aloha_3}) with the remaining wave function) which strongly limit the hope of further optimisation.

For the three gluon final state the situation is similar, even if the reduction of amplitude routine instructions is closer to a factor of 4. However, for this process the limiting factor is now the computation of the colour factor (taking around 60\% of the computation). We have also investigated how that step could be optimised and introduced  two simple improvements of the code. First we use a common sub-expression reduction algorithm \cite{Knuth:1976in} on that segment of the code. Second we merge the numerator and denominator of the colour-matrix into a single matrix, reducing the number of operations and allowing for a better memory pattern. Combined together those modifications lead to a speed-up of around 20\%.\footnote{One should note that the remaining contribution of the colour-matrix is auto-vectorisable and it is therefore advantageous to compile with hardware specific flag (giving an additional speed-up of around 10\%) (See how to do that in \ref{sec:manual}).}

\subsubsection{Overall speed-up} 
\label{sub:overall_speed_up}

\input{input/speedup}

Having looked at how the computational cost of \texttt{madevent} breaks down into different functions, we now present its overall factor speed increase. This is shown for a range of processes in table (\ref{tab: madevent}). If $t_{\text{with}}$ and $t_{\text{without}}$ are the times it takes to run \texttt{madevent} with and without helicity recycling respectively then the speed-up is represented as \[\text{speed-up} = \frac{t_{\text{without}}}{t_{\text{with}}}.\] 
As has already been alluded to in tables (\ref{tab: instr nohel}) and (\ref{tab: instr hel}), the speed-up we gain from helicity recycling is highly process dependent. Helicity recycling reduces the number of times we must calculate wave functions and amplitudes and so processes with more complicated diagrams and with a higher number of total diagrams see the biggest boost. For example, consider the $gg\rightarrow t \bar{t}$ results shown in table (\ref{tab: madevent ttbar}). As more gluons are added to the final state the gain increases, with the $t\bar{t}gg$ final state seeing a dramatic $2.27\times$ speed increase.

In contrast to this large increase the similar process $qq\rightarrow t \bar{t} q q$ (where $q \in \{u,d,c,s\}$) sees a noticeably lower speed increase of $1.27\times$. This is because processes involving fermions have a higher number of non-contributing helicity combinations and so helicity filtering will have a bigger effect. Hence, there will be fewer wave functions/amplitudes to evaluate and so helicity recycling will have a smaller impact.

One can see that the other processes presented in table (\ref{tab: madevent}) also follow these general principles regarding diagram complexity and fermion multiplicity. $W$ bosons allow for minimal helicity filtering and so table (\ref{tab: madevent wpwm}) displays a large speed increase, whereas electrons - being fermions - suppress the speed-up in table (\ref{tab: madevent e+e-}).

In table (\ref{tab: big hel comp}) we present these results for a wider range of processes and with a more detailed breakdown. In the `hel' column we present how many helicity combinations are evaluated per phase-space point. In this column we use a `$\to$' to indicate that the maximum value across the various matrix elements is being shown. The columns ``survey'' and ``refine'' present the timing of the two main phases of the phase-space integration/event generation for a single run. The ``survey''  is designed to get a first (un-precise) estimate of the cross section and to know the relative importance of each contribution. The amount of time spent in the ``survey'' is therefore independent of the number of requested events. The ``refine'' stage aims to generate the number of requested events and therefore scales linearly with the number of events.\footnote{The scaling is not perfectly linear due to various thresholds in the method of integration. For this table (and similar ones) we always request the code to generate ten thousand un-weighted events.} In both cases, the timing is the time to solution observed on a i7-7700HQ CPU (2016 macbook pro laptop) using up to 8 threads in the standard multi-core mode of the code when requesting ten thousand events.
\footnote{To first approximation, the precision on the cross section is directly related to the number of events generated, which means all integrals are estimated at the percent level.}
The last column presents the global speed-up (computed by comparing the sum of the timings of the survey and of the refine) between the version of the code including all the optimisations introduced in this paper (2.9.0) and the same code without helicity recycling (2.9.0 nohel). The optimisation related to the colour computation are present in both columns.
A detailed description of the cuts, scale choices and such are given as \href{https://doi.org/10.14428/DVN/B5NADE}{supplementary material}. There one can also find all the material required to reproduce this (and sub-sequent) tables.

Again one can see the same principles at play: the biggest speed increase is seen for complex processes dominated by QCD and vector boson interactions as they have the most non-vanishing helicity combinations and diagrams. The disparity of the gain between ``survey'' and ``refine'' can be explained due to different relative importances of the various matrix elements in each of the steps. 



Notice that helicity recycling's tendency to more greatly speed-up more intensive processes is very convenient. For example, in a $pp\rightarrow 4j$ simulation, the processes with $j=g$ will heavily dominate the calculation and so speeding those up is the most effective way to speed-up the overall simulation. This is why the simulation sees an impressive $2.1\times$ speed-up in table (\ref{tab: big hel comp}).

\begin{table*}[]
    \centering
    \begin{tabular}{l|c|c|c|c|c|c}
    process                                                                                   & hel           & \multicolumn{2}{c|}{2.9.0 nohel} &\multicolumn{2}{c|}{2.9.0} & speed-up   \\ \hline
    VBF-like processes                                                                        &               & survey & refine                  & survey & refine            &            \\
\hline 
 $p p \to W^+ W^+ j j\, [g_S=0]$ & 9 & 21s & 10m52s & 14s & 7m28s & 1.5$\times$ \\
 $p p \to W^+ W^- j j, W \to l vl [g_S=0$,13\,TeV] & $\to6$ & 10m0s & 2m5s & 7m0s & 1m29s & 1.4$\times$ \\
 $p p \to W^+ W^- j j, W \to l vl [g_S=0$,100\,TeV] & $\to6$ & 7m0s & 22m14s & 5m0s & 16m25s & 1.4$\times$ \\
 $u \bar{d} \to W^+_L W^-_L u \bar{d} [g_S=0]$ & 4 & 17s & 2m29s & 11s & 1m36s & 1.6$\times$ \\
  $u \bar{d} \to W^+_L W^-_L u \bar{d},  W^+ \to d \bar u, W^- \to \tau^+ \nu_\tau [g_S=0]$ & 8 & 1m0s & 8m19s & 1m0s & 4m27s & 1.7$\times$ \\
 $u \bar{d} \to W^+_T W^-_T u \bar{d}$,$ W^+ \to d \bar u, W^- \to \tau^+ \nu_\tau [g_S=0]$ & 8 & 40s & 2m16s & 33s & 2m6s & 1.1$\times$ \\
 $\mu^+ \mu^- \to h h h \bar\nu_\mu \nu_e$ [14\,TeV] & 1 & 1s & 11s & 1s & 10s & 1.1$\times$ \\
 $\mu^+ \mu^- \to t \bar{t} \mu^+ \mu^- $ [13\,TeV] & 64 & 23s & 1m35s & 5s & 22s & 4.4$\times$ \\
 $\mu^+ \mu^- \to W^+ W^- \mu^+ \mu^-$ [4\,TeV] & 122 & 1m0s & 46s & 14s & 14s & 3.8$\times$ \\
     \hline
     other processes & & survey & refine & survey & refine & \\
     \hline
 $p p \to W^+ [0-4]j$  & $\to48$ & 37m0s & 3s & 11m0s & 3s & 3.4$\times$ \\
 $p p \to t \bar{t} [0-2]j$  & $\to64$ & 1m0s & 49s & 27s & 22s & 2.2$\times$ \\
 $p p \to 4j$ & $\to50$ & 3m0s & 28m31s & 1m0s & 14m17s & 2.1$\times$ \\
 $p p \to t \bar{t} 3j$  & $\to128$ & 1h0m & 2h42m & 52m0s & 1h16m & 1.7$\times$ \\
 $p p \to W^+ Z$  & 9 & 1s & 3s & 1s & 2s & 1.3$\times$ \\
 $p p \to t \bar{t} h$ & $\to16$ & <1s & 3s & <1s & 2s & 1.5$\times$ \\
  $p p \to t \bar{t} h j $ & $\to32$ & 4s & 7s & 3s & 4s & 1.6$\times$ \\
 $p p \to t \bar{t} Z$ & $\to48$ & 1s & 9s & 1s & 4s & 2.0$\times$ \\
 $p p \to W^+ W^- j j$ [QCD only]  & $\to72$ & 40s & 1m17s & 10s & 33s & 2.7$\times$ \\

    \end{tabular}
    \caption{Integration timing (time to solution) on a mac-book pro (quad-core 2016) to generate 10k events. Comparing the impact with and without helicity recycling. 
    }
    \label{tab: big hel comp}
\end{table*}


\section{Phase-Space integrator}
\label{sec:ps}
\subsection{Monte-Carlo Integration and Single Diagram enhancement}
\label{sec:sdereview}
In addition to the speed of the matrix-element evaluation, the efficiency of the phase-space integrator is another crucial factor determining the speed of the computation since it controls how many times we need to evaluate the matrix element. Due to both the number of dimensions of integration and also the requirement to generate uncorrelated events, the only suitable method is Monte-Carlo integration. However, the convergence of the method is quite slow ($1/\sqrt{N}$, where $N$ is the number of times the matrix element is evaluated).

In Monte-Carlo methods (see \cite{Weinzierl:2000wd} for a nice review), one evaluates the integrand for random values of the variable of integration. The estimator of the integral ($I_N$) is simply given by the average of the function estimated across these  random points $x_i$:
\begin{equation}
\int f(x) \approx I_N = \frac1N \sum_{i=1}^N f(x_i).
\end{equation}
\noindent The statistical error ($\Delta I_N$) of this estimator $I_n$ is controlled by the variance of the integrand and can be estimated as
\begin{equation}
\Delta I_N = \frac{\text{Var}_f}{\sqrt{N}}\approx\frac{1}{\sqrt{N}} \sqrt{\frac1N\sum_{i=1}^N f^2(x_i) -I_N^2}. 
\end{equation}

Since the convergence rate is fixed to $\frac{1}{\sqrt{N}}$, the various avenues of optimisation consist in modifying the function being integrated to effectively reduce its variance. 
In \mg, we use the single diagram enhancement method \cite{Maltoni:2002qb}, which combines various importance sampling methods -- both analytical \cite{Weinzierl:2000wd} and numerical \cite{Lepage:2020tgj,Press:1989vk} -- on top of a multi-channel strategy. The integral is decomposed as a sum of various contributions:
\begin{eqnarray}
\int |M|^2 &=& \sum_i \int |M_i|^2 \frac{|M|^2}{\sum_j |M_j|^2},\label{eq:singlediagramenhancement}\\
        &\equiv&  \sum_i \int \alpha_i |M|^2,\label{eq:sde_alpha}
\end{eqnarray}
where the indices $i$ and $j$ range over individual (or potentially a subset of) Feynman diagrams. The values $\alpha_i\equiv \frac{|M_i|^2}{\sum_j |M_j|^2}$ are called the channel weights, they do not modify the value of the integral (as long as $\sum_i \alpha_i =1$ for every phase-space point) but do impact the variance of the integrand and therefore the speed of the computation.

While in general the value $\alpha_i$ could be any arbitrary function, the single diagram enhancement method makes a specific choice a-priori. This choice is particularly motivated by the classical limit where interference terms between the Feynman diagrams are small.
In that case
\begin{equation} \label{eq:PS-noint_approx}
\frac{|M|^2}{\sum_j |M_j|^2} \approx 1,
\end{equation}
\noindent and therefore we have 
\begin{equation}
\int |M|^2 =  \sum_i \int |M_i|^2 \frac{|M|^2}{\sum_j |M_j|^2} \approx \sum_i \int |M_i|^2.
\end{equation}
In other words, with this choice the assumption is that each of the terms of the sum -- called channels of integration -- are mainly behaving as a single diagram squared from which the poles are easily identifiable and importance sampling is relatively easy to implement
(see section (\ref{sec:tchannel})). The role of machine learning algorithms (\mg\ uses a modified VEGAS algorithm\cite{Lepage:2020tgj}) is then mainly to catch the impact of the $\frac{|M|^2}{\sum_j |M_j|^2}$ term as well as other sources of deviation from the ideal case (like for example the impact of generation cuts or the impact of the running of the strong coupling).

While this method works extremely well in general,  it is not the most suited approach for VBF-like processes, especially at high energy where large interference occurs due to gauge symmetry. As a matter of fact, \mgnlo\ is much slower than many other programs \cite{vbfnloconf} (in particular VBF\-NLO~\cite{Baglio:2014uba}) for such types of generation.

\subsection{$t$-channel strategy}
\label{sec:tchannel}

\begin{figure}
\begin{center}

  \includegraphics[width=0.4\textwidth]{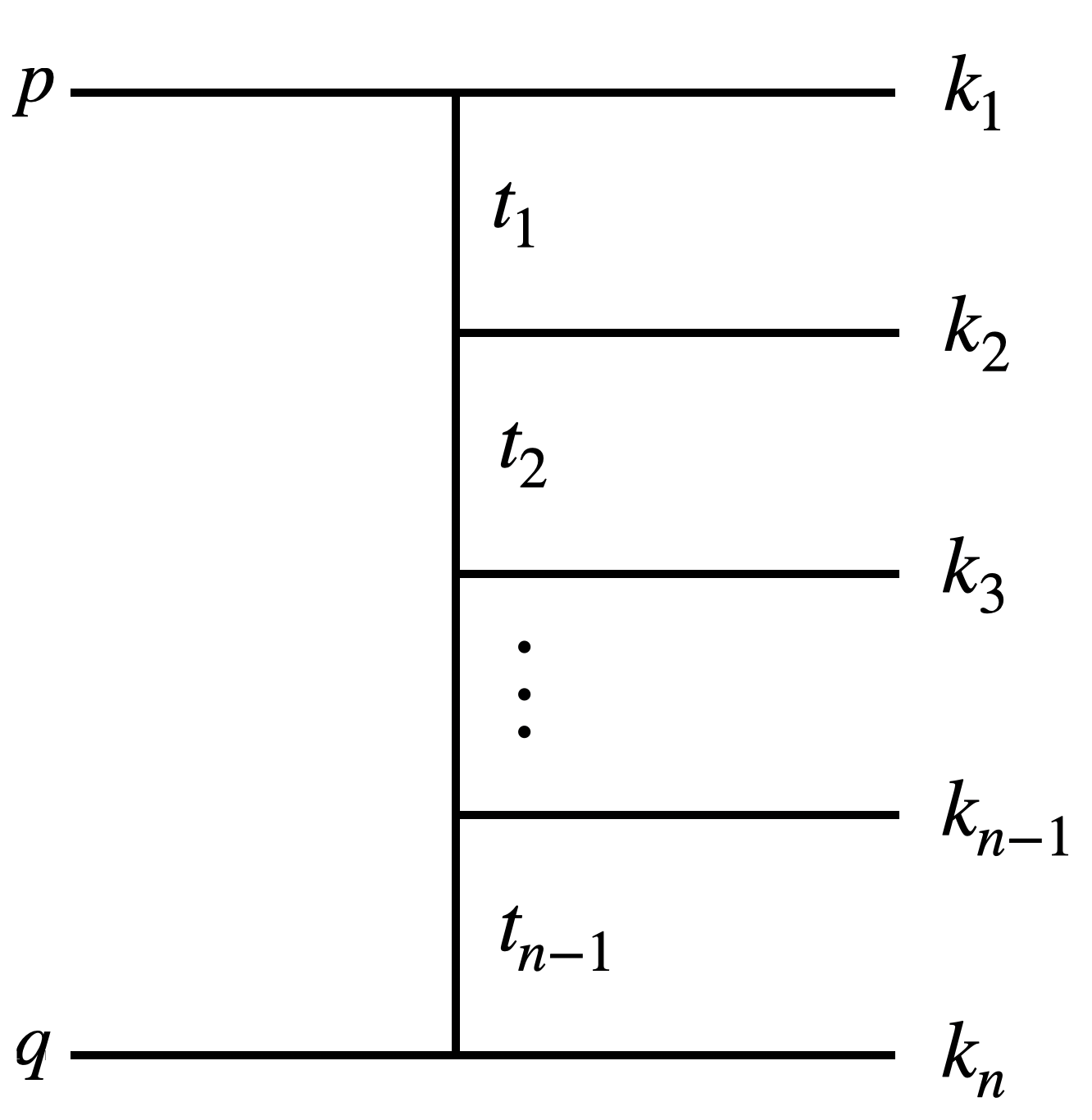}
  \end{center}
\caption{Feynman diagram used to fix the convention in the variable.}
\label{fig:ordering}       
\end{figure}

When running \mg, one can easily identify that the slowest channels of integration are the ones with multiple $t$-channel propagators.
In \mg, we handle $t$-channel propagators  with the following change of variable/phase-space measure \cite{Byckling:1971vca,Maltoni:2012pa}:
\begin{eqnarray}
    d\Phi_2 &=& \frac{d^3k_1}{(2\pi)^32E_1}\frac{d^3k_2}{(2\pi)^32E_2}
(2\pi)^4\delta^{(4)}(p+q-k_1-k_2),\\
   &=&\frac{1}{16\pi^2 \sqrt{\lambda(t_1,q^2,p_1^2)}} dt_1 d\phi,
\label{eq:tchannel}
\end{eqnarray}

\noindent where as in Fig. \ref{fig:ordering}, $p$ and $q$ are the initial state momenta and $k_1$ and $k_2$ the final state ones, $E_1$ (respectively $E_2$) is the energy of $k_1$ (respectively $k_2$), $S=(p+q)^2$ is the center of mass energy of the collision, $t_1$ is the virtuality of the $t$-channel propagator given by $t_1=(p-k_1)^2$ and where $\lambda(a,b,c)=a^2+b^2+c^2-2ab-2ac-2bc$.

The integration over $t_1$ is bounded by the following values:
\begin{eqnarray}
t^\pm_1&=& m_p^2+m_1^2 - \frac{1}{2S}\Big((S+m_p^2-m_q^2)(S+m_1^2-m_2^2)\nonumber\\
&&\mp \lambda^\frac12(S, m_p^2,m_q^2)\lambda^\frac12(S,m_1^2,m_2^2) \Big),
\label{eq:tbound}
\end{eqnarray}

\noindent In the presence of multiple $t$-channel propagators, \mg\ writes them as 
\begin{align}
t_1&=(p-k_1)^2
\nonumber\\
t_2&=(p-k_1-k_2)^2
\nonumber\\
&\ldots
\nonumber\\
t_{n-1}&=(p-k_1-k_2-\ldots-k_{n-1})^2
\end{align}
\noindent and performs the integration in the following (ordered) way
\begin{equation}
\int dt_{n-1}\int dt_{n-2} \dots \int dt_{1}, \label{eq:tordering}
\end{equation}
meaning that we first integrate over $t_{1}$, then $t_{2}$ and continues up to $t_{n-1}$.

\begin{figure*}
  \centering
  \begin{subfigure}{.35\textwidth}
    \centering
    \input{input/tg}
  \end{subfigure}%
  \begin{subfigure}{.65\textwidth}
    \centering
    \begin{tabular}{|c|c|c|c|}
    \hline
       order & unweighted events  &  relative precision & comment\\ \hline
       \multicolumn{4}{|c|}{3 iterations/20k evaluations}\\\hline
       $\int dt_3 \int dt_2 \int dt_1$  & 161 (.8\%) & 2.9\% & old default\\ 
       $\int dt_1 \int dt_2 \int dt_3$  & 1034 (5.0\%) & 1.4\% & new default\\
       $\int dt_2 \int dt_3 \int dt_1$ &  564 (2.8\%) & 2.2\% &\\
       $\int dt_2 \int dt_1 \int dt_3$ & 194 (1.0\%) & 2.8\% &\\ \hline 
        \multicolumn{4}{|c|}{5 iterations/ 80k evaluations}\\  \hline 
      $\int dt_3 \int dt_2 \int dt_1$   & 1703 (2.1\%)& 1.0\% & old default\\
      $\int dt_1 \int dt_2 \int dt_3$ & 5421  (6.8\%)& 0.6\% & new default\\
      $\int dt_2 \int dt_3 \int dt_1$ & 3636  (4.5\%)& 0.8\% &\\
      $\int dt_2 \int dt_1 \int dt_3$ & 4224 (5.3\%)& 0.8\% &\\

       \hline
    \end{tabular}
  \end{subfigure}
  \caption{Comparison of various orderings of the three variables of integration corresponding to the invariant of time-like particles for the channel associated to the Feynman diagram represented on the left. We present both the relative error, the number of events generated at a given iteration and the associated un-weighting efficiency.}
  \label{fig:ggttggordering}
\end{figure*}

The combination of such an ordering with the boundary condition of Eq. (\ref{eq:tbound}) creates correlations between the variables of integration. This correlation is problematic for any type of VEGAS-like algorithm \cite{Lepage:2020tgj,Weinzierl:2000wd}. The importance of the ordering can be seen in figure (\ref{fig:ggttggordering}) where we compare various orderings for a given channel of integration (the associated diagram is displayed in the figure). We present both the un-weighting efficiency and the estimated statistical uncertainty after three and five iterations which gives a hint on the convergence of the integration grid.

In this example, it is clear that the original ordering strategy picked by MG5aMC  is sub-optimal (between 3 and 6 times slower than the optimal strategy depending of the number of iterations). Not only is the best strategy more efficient at generating events but also the associated grid seems to need fewer iterations to converge.

While we hand-picked an example of channel of integration where the ordering was sub-optimal, it would not have been difficult to also present cases where it was optimal
(e.g. just flip the initial state in the example). Indeed the optimal ordering is deeply dependent on the channel of integration under consideration.

On top of the old ordering, we have added support for three additional orderings (see details in \ref{sec:manual}). Each channel of integration is then associated to one of those orderings
such that the virtuality of the most singular $t$-channel propagator is integrated first.
In principle one could use the values of $t_i^+$ (Eq. \ref{eq:tbound}) to choose such an ordering, but this is technically not possible and we used a simple heuristic approach for such determination.

\begin{table*}[]
    \centering
    \begin{tabular}{c|c|c|c|c|c}
    process &\multicolumn{2}{|c|}{2.9.0 old ordering} &\multicolumn{2}{|c|}{2.9.0} & speed-up  \\ \hline
    VBF-like processes & survey & refine & survey & refine  \\
    \hline
 $p p \to W^+ W^+ j j\, [g_S=0]$  & 17s & 7m1s & 14s & 7m28s & {\color{red} 0.91$\times$} \\
 $p p \to W^+ W^- j j, W \to l vl [g_S=0$,13\,TeV]  & 8m0s & 1m30s & 7m0s & 1m29s & 1.1$\times$ \\
 $p p \to W^+ W^- j j, W \to l vl [g_S=0$,100\,TeV]  & 6m0s & {\bf 1h0m}/3821 & 5m0s & 16m25s & 3.1$\times$ \\
 $u \bar{d} \to W^+_L W^-_L u \bar{d} [g_S=0]$  & 12s & 4m23s & 11s & 1m36s & 2.6$\times$ \\
  $u \bar{d} \to W^+_L W^-_L u \bar{d},  W^+ \to d \bar u, W^- \to \tau^+ \nu_\tau [g_S=0]$  & 1m0s & 24m19s & 1m0s & 4m27s & 4.6$\times$ \\
 $u \bar{d} \to W^+_T W^-_T u \bar{d}$,$ W^+ \to d \bar u, W^- \to \tau^+ \nu_\tau [g_S=0]$  & 33s & 1m9s & 33s & 2m6s & {\color{red} 0.63$\times$} \\
 $\mu^+ \mu^- \to h h h \bar\nu_\mu \nu_e$ [14\,TeV]  & 1s & 29s & 1s & 10s & 2.7$\times$ \\
 $\mu^+ \mu^- \to t \bar{t} \mu^+ \mu^- $ [13\,TeV]  & 7s & 24s & 5s & 22s & 1.1$\times$ \\
 $\mu^+ \mu^- \to W^+ W^- \mu^+ \mu^-$ [4\,TeV]  & 13s & 15s & 14s & 14s & 1.0$\times$ \\
     \hline
     other processes & survey & refine & survey & refine & \\
     \hline
 $p p \to W^+ [0-4]j$   & 14m0s & 3s & 11m0s & 3s & 1.3$\times$ \\
 $p p \to t \bar{t} [0-2]j$   & 36s & 27s & 27s & 22s & 1.3$\times$ \\
 $p p \to 4j$  & 1m0s & 15m26s & 1m0s & 14m17s & 1.1$\times$ \\
 $p p \to t \bar{t} 3j$   & 1h0m & 3h44m & 52m0s & 1h16m & 2.2$\times$ \\
 $p p \to W^+ Z$   & 1s & 2s & 1s & 2s & 1.0$\times$ \\
 $p p \to t \bar{t} h$  & <1s & 1s & <1s & 2s & {\color{red} 0.5$\times$} \\
  $p p \to t \bar{t} h j $  & 2s & 5s & 3s & 4s & 1.0$\times$ \\
 $p p \to t \bar{t} Z$  & 1s & 4s & 1s & 4s & 1.0$\times$ \\
 $p p \to W^+ W^- j j$ [QCD only]   & 9s & 1m24s & 10s & 33s & 2.2$\times$ \\
 \end{tabular}
    \caption{Integration timing (time to solution) on a mac-book pro (quad-core 2016) to generate 10k events depending of the $t$-channel ordering strategy. If the generation fails to generate enough events the timing is re-scaled accordingly. In such a case the timing is set in bold and the number of events actually generated is then indicated.}
    \label{tab:ps_tstrategy}

\end{table*}

Our resultant timings are presented in table (\ref{tab:ps_tstrategy}), which contains the same type of information as table (\ref{tab: big hel comp}). The comparison is between the version of the code containing all the optimisations of this paper (2.9.0) with the same version of the code where the ordering strategy was forced to the old one  (Eq. \ref{eq:tordering}).
While in general the ``survey'' presents only a small speed-up \footnote{This is due to the requirement of the code to perform at least three iterations.} a more sizeable gain is achieved during the ``refine''. The actual speed-up will therefore slightly increase when requesting more events.
Additionally, one can note that the biggest gain is achieved for the slowest processes.

One process (VBF with transverse polarised W boson) shows a significant slow-down (1.6 times slower).
Our investigation shows that the ordering picked in that case was correct but the convergence of the grid was actually slower than what it was for the previous strategy leading to the slow-down. Such an issue can only happen for relatively easy channels of integration since most complex processes would need more iterations and then this effect would disappear (as observed on the VBF process at 100 TeV).

\subsection{New Diagram enhancement strategy}
\label{sec:propa}

A recent paper \cite{Hagiwara:2020tbx} pointed to the importance of the gauge choice when generating events within \mgnlo. Even if the amplitude is fully gauge invariant, the definition of the channel weights ($\alpha_i$) is not. The presence of large gauge cancellations  are related to large interference terms and therefore the assumptions used by the phase-space integrator  (Eq.  (\ref{eq:PS-noint_approx})) will not hold anymore. Consequently, the $\alpha_i$ associated with Feynman diagrams with $t$-channel propagators will be artificially large at high-energy. This will increase the variance of the function and reduce the efficiency of the method.

We propose here, a second strategy for the single diagram enhancement method. Instead of using Eq. (\ref{eq:singlediagramenhancement}), we replace the $|M_i|^2$ by the product of the denominator (both $s$ and $t$ channel) associated to the single diagram under consideration (and normalise them as needed).
\begin{eqnarray}
    \bar \alpha_i &\equiv&  \prod_{\substack{k \in \\\text{propagator}}} \frac{1}{|p_k^2-M_k^2 -iM_k\Gamma_k|^2},\\
    \alpha_i &=& \frac{\bar \alpha_i}{\sum_j \bar \alpha_j}. \label{eq:sdenew-wgt} 
\end{eqnarray}

Such a change in the definition of the multi-weights does not in principle impact the cross section (see Eq. \ref{eq:sde_alpha}). However in practise the choice of the dynamical scale is done on a  CKKW inspired clustering \cite{Catani:2001cc} which depends on the Feynman diagram selected by the single diagram enhancement method. 
This modification of the channel weight will therefore impact the scale associated to each event and therefore the cross section and shape, both within scale uncertainties.\footnote{In our tests, the impact of this effect was at the order of the percent, so well below the scale uncertainty.}
In general this can be avoided --if needed-- by changing the scale computation to $H_T/2$ or some other purely kinematical scale choice \cite{Hirschi:2015iia}. However, one should note that it is not possible to avoid such effects when running matched/merged generation within MLM or shower-kT MLM \cite{Caravaglios:1998yr,Alwall:2008qv}, since the  CKKW clustering is mandatory in those type of generation.

A more subtle effect of the modification of the channel weight is related to the parton shower. When writing an event inside the output file, \mgnlo\ provides (in addition to the kinematic variables) the colour dipole representation in the leading colour approximation. The determination of the eligible dipole depends on the Feynman diagram (e.g. in presence of mixed expansion) and therefore the modification of the multi-channel strategy can impact such a selection. One can therefore expect some change, within theoretical uncertainties, after parton shower for some  QCD related observable. 

\begin{table*}[]
    \centering
    \begin{tabular}{c|c|c|c|c|c|c}
    process &\multicolumn{2}{|c|}{old strategy} &\multicolumn{2}{|c|}{new strategy} & speed-up & default \\ \hline
    VBF-like processes & survey & refine & survey & refine & \\
    \hline
 $p p \to W^+ W^+ j j\, [g_S=0]$  & 13s & {\bf 2h12m}/1290 & 16s & 8m1s & 16$\times$ & new\\
 $p p \to W^+ W^- j j, W \to l vl [g_S=0$,13\,TeV]  & 19m0s & 9m6s & 10m0s & 1m43s & 2.4$\times$ & new\\
 $p p \to W^+ W^- j j, W \to l vl [g_S=0$,100\,TeV]  & 10m0s & 24m8s & 7m0s & 18m10s & 1.4$\times$ & new\\
 $u \bar{d} \to W^+_L W^-_L u \bar{d} [g_S=0]$  & 23s & {\bf 27h56m}/203 & 14s & 1m53s & 792$\times$ & new\\
  $u \bar{d} \to W^+_L W^-_L u \bar{d},  W^+ \to d \bar u, W^- \to \tau^+ \nu_\tau [g_S=0]$  & 2m0s & {\bf 15h52m}/793 & 1m0s & 5m42s & 142$\times$ & new\\
 $u \bar{d} \to W^+_T W^-_T u \bar{d}$,$ W^+ \to d \bar u, W^- \to \tau^+ \nu_\tau [g_S=0]$  & 36s & 2m54s & 37s & 2m28s & 1.1$\times$ & new\\
 $\mu^+ \mu^- \to h h h \bar\nu_\mu \nu_e$ [14\,TeV]  & 3s & {\bf 8h50m}/641 & 1s & 11s & 2653$\times$ & new\\
 $\mu^+ \mu^- \to t \bar{t} \mu^+ \mu^- $ [13\,TeV]  & 20s & {\bf 3h6m}/948 & 6s & 25s & 362$\times$ & new\\
 $\mu^+ \mu^- \to W^+ W^- \mu^+ \mu^-$ [4\,TeV]  & 1m0s & 33m26s & 16s & 15s & 66$\times$ & new\\
     \hline
     other processes & survey & refine & survey & refine & \\
     \hline
 $p p \to W^+ [0-4]j$   & 20m0s & 5s & 20m0s & 4s & 1.0$\times$ & old\\
 $p p \to t \bar{t} [0-2]j$   & 38s & 32s & 38s & 19s & 1.2$\times$ & old\\
 $p p \to 4j$  & 1m0s & {\bf 1h21m}/7003 & 1m0s & 21m5s & 3.7$\times$ & new\\
 $p p \to t \bar{t} 3j$   & 1h0m & 1h36m & 2h0m & 1h37m & {\color{red} 0.71$\times$} & old\\
 $p p \to W^+ Z$   & 1s & 3s & 1s & 2s & 1.3$\times$ & new\\
 $p p \to t \bar{t} h$  & <1s & 2s & <1s & 3s & {\color{red} 0.67$\times$} & old\\
  $p p \to t \bar{t} h j $  & 2s & 4s & 3s & 10s & {\color{red} 0.45$\times$} & old\\
 $p p \to t \bar{t} Z$  & 1s & 4s & 1s & 4s & 1.0$\times$ & old\\
 $p p \to W^+ W^- j j$ [QCD only]   & 11s & 36s & 11s & 37s & 1.0$\times$ & old\\ 
    \end{tabular}
    \caption{Integration timing (time to solution) on a mac-book pro (quad-core 2016) to generate 10k events depending of the multi-channel strategy. If the generation fails to generate enough events the timing is re-scaled accordingly. In such a case the timing is set in bold and the number of events actually generated is then indicated.}
    \label{tab:ps_sde}
\end{table*}

In table (\ref{tab:ps_sde}), we compare the time needed to generate ten thousand events with the previous strategy and the one introduced in this paper (all other optimisations of this paper are included in both cases).
As one can expect for such deep modification of the phase-space integration strategy,
the observed spectrum of speed-up/down is extremely broad going from three orders of magnitude speed-up to five times slower. 
It is clear that such an optimisation is a must-have in the context of VBF processes but must be avoided for most QCD multi-jet processes. While the user can easily switch from one strategy to the other (see \ref{sec:manual}), we have configured the code such that the default value is process dependent. All processes with only one colour-flow will use the new method while others processes will use the old method. We made an exception for pure multi-jet processes which now use the new method as well.
The default for each process is indicated in the last column in table (\ref{tab:ps_sde}). 
Since for most QCD processes we keep the previous integration strategy, the caveats on the change in the scale/leading colour choice, mentioned above, are naturally mitigated.

\subsection{Comparison with older version of \mgnlo}
\label{sec:psresult}

\begin{table*}[]
    \centering
    \begin{tabular}{c|c|c|c|c|c|c}
    process &\multicolumn{2}{|c|}{2.8.1} &\multicolumn{2}{|c|}{2.9.0} & speed-up & x-section \\ \hline
    VBF-like processes & survey & refine & survey & refine & &(pb)\\
    \hline
 $p p \to W^+ W^+ j j\, [g_S=0]$  & 15s & 16m40s & 14s & 7m28s & 2.2$\times$ & 0.2\\
 $p p \to W^+ W^- j j, W \to l vl [g_S=0$,13\,TeV]  & 18m0s & 22m54s & 7m0s & 1m29s & 4.8$\times$ & 0.018\\
 $p p \to W^+ W^- j j, W \to l vl [g_S=0$,100\,TeV]  & 8m0s & {\bf 11h19m}/1398 & 5m0s & 16m25s & 32$\times$ & 0.66\\
 $u \bar{d} \to W^+_L W^-_L u \bar{d} [g_S=0]$  & 21s & {\bf 3h53m}/1497 & 11s & 1m36s & 131$\times$ & 0.00029\\
  $u \bar{d} \to W^+_L W^-_L u \bar{d},  W^+ \to d \bar u, W^- \to \tau^+ \nu_\tau [g_S=0]$  & 1m0s & {\bf 4h55m}/3120 & 1m0s & 4m27s & 54$\times$ & 1.2e-05\\
 $u \bar{d} \to W^+_T W^-_T u \bar{d}$,$ W^+ \to d \bar u, W^- \to \tau^+ \nu_\tau [g_S=0]$  & 43s & 8m36s & 33s & 2m6s & 3.5$\times$ & 9e-05\\
 $\mu^+ \mu^- \to h h h \bar\nu_\mu \nu_e$ [14\,TeV]  & 2s & {\bf 106h0m}/43 & 1s & 10s & 34693$\times$ & 6.9e-06\\
 $\mu^+ \mu^- \to t \bar{t} \mu^+ \mu^- $ [13\,TeV]  & 53s & {\bf 1h25m}/6401 & 5s & 22s & 190$\times$ & 0.033\\
 $\mu^+ \mu^- \to W^+ W^- \mu^+ \mu^-$ [4\,TeV]  & 10m0s & 1h55m & 14s & 14s & 267$\times$ & 2.4\\
     \hline
     other processes & survey & refine & survey & refine & \\
     \hline
 $p p \to W^+ [0-4]j$   & 38m17s & 4s & 11m0s & 3s & 3.5$\times$ & 1.1e+05\\
 $p p \to t \bar{t} [0-2]j$   & 1m32s & 1m27s & 27s & 22s & 3.7$\times$ & 1.5e+03\\
 $p p \to 4j$  & 3m26s & 4h18m & 1m0s & 14m17s & 17$\times$ & 1.3e+07\\
 $p p \to t \bar{t} 3j$   & 2h15m & 5h42m & 52m0s & 1h16m & 3.7$\times$ & 1.8e+02\\
 $p p \to W^+ Z$   & 1s & 8s & 1s & 2s & 3.0$\times$ & 15\\
 $p p \to t \bar{t} h$  & 1s & 3s & <1s & 2s & 2.0$\times$ & 0.38\\
  $p p \to t \bar{t} h j $  & 4s & 14s & 3s & 4s & 2.6$\times$ & 0.46\\
 $p p \to t \bar{t} Z$  & 1s & 7s & 1s & 4s & 1.6$\times$ & 0.56\\
 $p p \to W^+ W^- j j$ [QCD only]   & 41s & 3m13s & 10s & 33s & 5.4$\times$ & 23\\
    \end{tabular}
    \caption{Integration timing (time to solution) on a mac-book pro (quad-core 2016) to generate 10k events. If in 2.8.1, it was not possible to generate ten thousand events, the time is re-scaled accordingly (and the timing is set in bold). The second number presented in that case is the actual number of events that was generated.}
    \label{tab:ps_2.8.1}
\end{table*}

In table (\ref{tab:ps_2.8.1}), we compare the speed of the code between two versions of \mgnlo\ (2.8.1 and 2.9.0). 2.9.0 is the first version of the code containing the modification described in this paper. Let's stress that every optimisation flag is set to their default value and therefore this is the speed-up that a user will observe without playing with any options. 

The combined impact of all our optimisations is striking for VBF-like processes with a speed-up of more than 30,000 times faster for one process. While this process is very specific and probably not the most important one for many users, all the  VBF processes show massive speed-up passing from hour long runs to a couple of minutes. Actually, in many cases, the previous version of the code had a lot of trouble in generating the requested number of events and sometimes even to correctly converge to the correct cross section. All those problems are now solved with 2.9.0; the cross section converges quickly and events are generated very efficiently.

The gain for the other processes is more modest. Firstly because the phase-space integration was much better handled to start with and secondly because those processes are less sensitive to $t$-channel diagrams on which we have focused. Nevertheless in combination with the helicity recycling, the code is, for processes heavily used at the LHC, around three times faster, a very valuable gain.


\section{Conclusion}
\label{sec:conclusion}
In order to evaluate an amplitude \mg\ must sum it over all contributing helicity combinations. Before the work of this paper \mg\ would calculate every wave function and amplitude separately for each helicity combination. We have successfully restructured the code so it will now only calculate a given wave function once and then reuse the output for all the different helicity combinations. We have also been able to split up the amplitude calculation such that part of it can be recycled across different helicity combinations. This restructuring of the code has also allowed us to avoid calculating parts of the helicity calculation that contribute to null-diagrams. All these optimisations mean that for complex processes with few fermions we can a see a speed-up of the code of around $2\times$. At the other end of scale, simple processes dominated by fermions can see a speed-up of only a few percent.

Additionally, we have studied the efficiency issue of \mgnlo\ for VBF-like processes at high energy. We have identified that modifying the order of integration of the virtuality of $t$-channel particles and changing the multi-channel weight was highly valuable, providing game-changing speed-up for such computations. This has fixed a lot of the issues faced in the past for such processes.

Combining all those optimisations allow us to overshoot the target speed-up asked by the HSF community \cite{Aarrestad:2020ngo,Alves:2017she} since we provide a code at least three times faster for CPU intensive processes and far higher for VBF processes.

Optimisation is important for any program heavily used by LHC experiments and \mgnlo represents a non-negligible amount of grid/local cluster usage. We believe that this paper is a significant milestone for \mgnlo\, providing significant speed improvements both in the time to evaluate a phase-space point and to the phase-space integrator. However this is certainly not the end of the road and this effort has to (and will) continue. First, the techniques developed in this paper need to be applied to the NLO processes within \mgnlo. We do not expect any big difficulties in such porting and expect similar gain in speed. 
Second, there is still room to optimise the evaluation of the matrix element, even at leading order: work is in progress to have a tighter link to the hardware with investigation on a GPU port but also on the use of SIMD operation on CPUs \cite{Reinders2020}.

\appendix

\section{Manual}
\label{sec:manual}

\begin{table*}[]
    \centering
    \begin{tabular}{|c|c|c|l|}
    \hline
    \multicolumn{4}{|c|}{Options controlling helicity recycling} \\
    \hline
    parameter & default  & values & comments \\
     \hline
    hel\_recycling & T & T/F & allows/forbids the full helicity recycling \\
    hel\_filtering & T & T/F & allows filtering over non contributing helicity \\
    hel\_splitamp & T & T/F & allows amplitude splitting (See Eq. (\ref{aloha_1}-\ref{aloha_3})) \\
    hel\_ampzero & T & T/F & allows filtering of vanishing amplitudes \\
     \hline
      \hline
     \multicolumn{4}{|c|}{Options controlling phase-space integration} \\
         \hline
    parameter & default  & values & comments \\
     \hline
     sde\_strategy &  & 1/2 & integration strategy: ``1'' means Eq. (\ref{eq:singlediagramenhancement}), ``2'' means Eq. (\ref{eq:sdenew-wgt})\\
     hard\_survey & 0 & 0/1/2/3 & increase number of events and maximum number of iterations \\
     second\_refine\_threshold & 0.9 & [0,1] & threshold used to forbid the second refine \\
      \hline
      \multicolumn{4}{|c|}{Options controlling advanced compilation flag} \\
      \hline
      global\_fflag & "-O" & string & compilation flag for all used for all compilation \\
      aloha\_fflag & " " & string & additional compilation flag for aloha routine (suggested: -ffast-math) \\
       \multicolumn{4}{|r|}{Note: re-compilation is not automatic (need ``make clean'' in ``Source'' directory)}\\
      \hline

    \end{tabular}
    \caption{Short description of the various hidden parameters that can be specified within the run\_card (configuration file) to tweak the behaviour of the new methods introduced in this paper.}
    \label{tab:runcard}
\end{table*}

When generating an LO process within \mgnlo, the optimisation described in this paper 
will be activated by default (since version 2.9.0). If for some reason one wants to de-activate some optimisation or change some internal parameter, we offer the possibility to do so via a couple of parameters that can be included in the \texttt{run\_card.dat} which is the main configuration file. Most of these parameters 
are not present by default in that card since we consider them as advanced parameters. It is enough to add them in the file when needed. The only parameter present by default is the one allowing for the choice of multi-channel strategy.

In table (\ref{tab:runcard}), we briefly present the various parameters that can be modified and in what way. One should notice that the parameter ``hel\_recycling'' is a switch that forbids the use of helicity recycling, therefore when set to False, the other parameters related to helicity recycling (prefixed with ``hel\_'') will be without any impact.

It is also possible to ask directly \mgnlo\ to generate code without any helicity recycling. This allows for the generation of code closer to the previous version and avoids spending time generating the new aloha routines. To do this one needs to modify the ``output'' command and add the flag \\
``\texttt{-{}-hel\_recycling=False}''. For example
\begin{lstlisting}[language=Bash]
    generate p p > t t~ 3j
    output MYDIR --hel_recycling=False
\end{lstlisting}
Another new optional flag for the ``output'' command allows for the control of the ordering of the variable of integration corresponding to the invariant mass of $t$-channel propagators.
For example:
\begin{lstlisting}[language=Bash]
    generate p p > W+ W- j j QCD=0
    output MYDIR --t_strategy=X
\end{lstlisting}
The possible values and the associated meaning is described below. 
A concrete example for the ordering of the Feynman diagram represented in Fig. (\ref{fig:ggttggordering}) is also given to provide a simpler comparison.

\begin{itemize}
    \item {\bf 0} [default] Automatically decide based on the diagram.
    \item {\bf 1} Always use one-sided ordering integrating the $t$ invariant mass from the bottom of the Feynman Diagram to the top (initial state particle with positive $p_z$ is displayed conventionally at the top of the diagram). For the example, this corresponds to $\int dt_1 \int dt_2 \int dt_3$.
    \item {\bf 2} Always use one-sided ordering integrating the $t$ invariant mass from the top of the Feynman Diagram to the bottom. This was the only option in older versions of the code. For the example, this correspond to  $\int dt_3 \int dt_2 \int dt_1$.
    \item {\bf -1} Going from external to internal $t$-variables starting with the $t$ invariant mass at the bottom of the Feynman Diagram, then the one at the top, then the second one from the bottom, followed by the second one from the top and so on up to depletion. For the example, this correspond to $\int dt_2 \int dt_1 \int dt_3$.
    For an example with 4 t-channel propagator this will correspond to 
    $\int dt_2 \int dt_3 \int dt_1 \int dt_4$.
    \item {\bf -2} Same as the previous ordering but starting from the most top invariant mass. For the example, this correspond to $\int dt_2 \int dt_3 \int dt_1$ and for a 4 t-channel propagator case, this will correspond to $\int dt_3 \int dt_2 \int dt_4 \int dt_1$.
\end{itemize}

A final flag for the ``output'' command allows for the deactivation of the common sub-expression reduction algorithm for the colour-factor part of the code (which by default is activated):
\begin{lstlisting}[language=Bash]
    generate g g > t t~ g g g
    output MYDIR --jamp_optim=False
\end{lstlisting}
In some cases this flag can be relevant since the upfront cost of such a function can be large (here 5 min for the example), while the run-time gain (around 8\%) might not always be enough to justify it. In general both the upfront cost and the reduction are barely noticeable. Note that this flag can also be used for Fortran standalone output.

\section{Extension of {\sc Aloha}}
\label{sec:aloha}

\aloha\ \cite{deAquino:2011ub} is a program shipped with \mgnlo\ which computes automatically the set of helicity amplitudes needed for a given matrix-element computation (like Eq. \ref{eq:photon_wfct}). After the generation of the Feynman diagram by \mgnlo, \mgnlo\ requests \aloha\ to generate the HELAS \cite{Murayama:1992gi} function needed for the associated computation.

In the context of helicity recycling a new type of helicity routine has been introduced (Eq. (\ref{aloha_1}-\ref{aloha_2})). Contrary to the other types of routines, \mgnlo\
does not know at generation time which of those functions will be effectively used.
 Consequently \mgnlo\ request \aloha\ to generate all possible choices (so in general three routines) such that any strategy can be picked if relevant. 

The implementation strategy is to ask \aloha\ to generates a standard internal wave-function routine but with a custom propagator. This was possible thanks to a previous extension of \aloha\  adding support for custom propagators \cite{Christensen:2013aua,BuarqueFranzosi:2019boy}.
The definitions for such propagators are 
\begin{eqnarray}
\textrm{scalar}&:\:\: & 1, \\
\textrm{fermion}&:\:\: & \delta_ij,\\
\textrm{vector}&:\:\: & \eta^{\mu\nu}.
\end{eqnarray}

The reason for the presence of a metric term for the vector propagator is that it allows us to not include the metric in the final scalar product and therefore have code which is easier for the compiler to optimise (giving the possibility to use some SIMD instructions) which can be critical since a large part of the computation is spent evaluating such simple scalar products (\(\approx\) 20\% for $g g \to t\bar{t} g g$).

Concerning the ALOHA naming scheme convention, such a routine will have a suffix ``P1N''.
So for the following Lorentz structure (which correspond to a $\gamma_\mu$ Lorentz structure \cite{Degrande:2011ua}):
\begin{lstlisting}[language=Bash]
FFV1 = Lorentz(name = 'FFV1',
    spins = [ 2, 2, 3 ],
    structure = 'Gamma(3,2,1)')
\end{lstlisting}
the three new expressions (\ref{aloha_1}-\ref{aloha_2}) will have the following name/definitions:
\begin{eqnarray}
\textrm{FFV1P1N\_1} & : & \gamma_\mu\, \psi^{h_2}_2 \phi^\mu_{h_\phi},\\ 
\textrm{FFV1P1N\_2} & : & \bar{\psi}^{h_1}_1\, \gamma_\mu\, \phi^\mu_{h_\phi},\\
\textrm{FFV1P1N\_3} & : & \bar{\psi}^{h_1}_1\, \gamma_\mu\, \psi^{h_2}_2 \eta^{\mu\nu}.
\end{eqnarray}

%
%

\begin{acknowledgements}
The authors would like to thanks Fabio Maltoni, Mike Seymour, Richard Ruiz, Luca Mantani, Andrew Lifson, Andrea Valassi, Stefan Roiser and all \mgnlo\ authors (past and present) for useful discussions. We would also like to thank the Université catholique de Louvain staff for working around the limitations imposed by the Covid-19 pandemic. This work has received funding from the European Union's Horizon 2020 research and innovation programme as part of the Marie Skłodowska-Curie Innovative Training Network MCnetITN3 (grant agreement no. 722104). This project has received funding from the European Union’s Horizon 2020 research and innovation programme under grant agreement No 824093.
Computational resources have been provided by the Consortium des Équipements de Calcul Intensif (CÉCI), funded by the Fonds de la Recherche Scientifique de Belgique (F.R.S.-FNRS) under Grant No. 2.5020.11 and by the Walloon Region.

\end{acknowledgements}

\bibliographystyle{spmpsci}      
\bibliography{paper.bib}   

%
%

\end{document}

%% file: input/hel_amp.tex
\begin{tikzpicture}
	\begin{feynman}
		\vertex (i1);
		\vertex [below right=1.5cm of i1] (a);
		\vertex [below right=1cm of i1] (a1);
		\vertex [below left=1.5cm of a] (i2);
		\vertex [below left=.5cm of a] (a2);
		\vertex [right=1cm of a] (b);
		\vertex [above right=0.5cm of b] (b2);
		\vertex [above right=0.75cm of b] (x);
		\vertex [below right=0.2cm of x] (x1) ;
		\vertex [above right=0.2cm of x] (x2) ;
		\vertex [right=1.2cm of b] (f3) ;
		\vertex [above right=1.5cm of b] (f1) ;
		\vertex [below right=.5cm of b] (b1);
		\vertex [below right=1.5cm of b] (f2) ;

		\vertex [below right=1.3cm and .2cm of a] (cap) {(a)};

		\diagram* {
		(i1) -- [fermion] (a1),
		(a2) -- [fermion] (i2),
		(f2) -- [fermion] (b1),
		(x2) -- [fermion] (f1),
		(x1) -- [gluon] (f3),
		};
	\end{feynman}

	\node [right=0.2cm of f3] (ar1) {};
	\node [right=0.7 of ar1] (ar2) {};
	\node [right=0.2 of ar1] (ar_centre) {};

	\node [above=.8 of ar_centre] (up) {};

	\node [below=.8 of ar_centre] (down) {};

	\draw [->, ultra thick] (ar1) -- (ar2);

\end{tikzpicture}%
\begin{tikzpicture}
	\begin{feynman}
		\vertex (i1);
		\vertex [below right=1.5cm of i1] (a);
		\vertex [below right=1cm of i1] (a1);
		\vertex [below left=1.5cm of a] (i2);
		\vertex [below left=.5cm of a] (a2);
		\vertex [right=1cm of a] (b);
		\vertex [above right=0.25cm of b] (b2);
		\vertex [above right=0.75cm of b] (x);
		\vertex [right=1.2cm of b] (f3) ;
		\vertex [above right=1.5cm of b] (f1) ;
		\vertex [below right=.5cm of b] (b1);
		\vertex [below right=1.5cm of b] (f2) ;

		\vertex [below right=1.3cm and .2cm of a] (cap) {(b)};

		\diagram* {
		(i1) -- [fermion] (a1),
		(a2) -- [fermion] (i2),
		(f2) -- [fermion] (b1),
		(b2) -- [red] (x) -- [fermion] (f1),
		(x) -- [gluon] (f3),
		};

	\end{feynman}
	 \draw[color=green!40!black!60!, ultra thick] (x) circle (4pt);

	\node [right=0.2cm of f3] (ar1) {};
	\node [right=0.7 of ar1] (ar2) {};
	\node [right=0.2 of ar1] (ar_centre) {};

	\node [above=.8 of ar_centre] (up) {};

	\node [below=.8 of ar_centre] (down) {};

	\draw [->, ultra thick] (ar1) -- (ar2);

\end{tikzpicture}
\begin{tikzpicture}
	\begin{feynman}
		\vertex (i1);
		\vertex [below right=1.5cm of i1] (a);
		\vertex [below left=1.5cm of a] (i2);
		\vertex [right=1cm of a] (b);
		\vertex [above right=0.25cm of b] (b2);
		\vertex [above right=0.75cm of b] (x);
		\vertex [right=1.2cm of b] (f3) ;
		\vertex [above right=1.5cm of b] (f1) ;
		\vertex [below right=.5cm of b] (b1);
		\vertex [below right=1.5cm of b] (f2) ;

		\vertex [below right=1.3cm and .2cm of a] (cap) {(c)};

		\diagram* {
		(i1) -- [fermion] (a)-- [fermion] (i2),
		(a) -- [red, photon] (b),
		(f2) -- [fermion] (b1),
		(b2) -- (x) -- [fermion] (f1),
		(x) -- [gluon] (f3),
		};

	\end{feynman}
	\draw[color=green!40!black!60!, ultra thick] (a) circle (4pt);

	\node [right=0.2cm of f3] (ar1) {};
	\node [right=0.7 of ar1] (ar2) {};
	\node [right=0.2 of ar1] (ar_centre) {};

	\node [above=.8 of ar_centre] (up) {};

	\node [below=.8 of ar_centre] (down) {};

	\draw [->, ultra thick] (ar1) -- (ar2);

\end{tikzpicture}%
\begin{tikzpicture}
	\begin{feynman}
		\vertex (i1);
		\vertex [below right=1.5cm of i1] (a);
		\vertex [below left=1.5cm of a] (i2);
		\vertex [right=1cm of a] (b);
		\vertex [above right=0.75cm of b] (x);
		\vertex [right=1.2cm of b] (f3) ;
		\vertex [above right=1.5cm of b] (f1) ;
		\vertex [below right=1.5cm of b] (f2) ;

		\vertex [below right=1.3cm and .2cm of a] (cap) {(d)};

		\diagram* {
		(i1) -- [fermion] (a)-- [fermion] (i2),
		(a) -- [photon] (b),
		(f2) -- [fermion] (b) -- (x) -- [fermion] (f1),
		(x) -- [gluon] (f3),
		};

	\end{feynman}
	\draw[color=green!40!black!60!, ultra thick] (b) circle (4pt);
\end{tikzpicture}

%% file: input/instr_nohel.tex
\begin{table}
    \renewcommand{\arraystretch}{1.5}
    \centering
    \begin{tabular}{ l | r@{}@{}l | r@{}@{}l | r@{}@{}l}
        & \multicolumn{2}{c|}{\(gg\rightarrow t\bar{t}\)} & \multicolumn{2}{c|}{\(gg\rightarrow t\bar{t}gg\)} & \multicolumn{2}{c}{\(gg\rightarrow t\bar{t}ggg\)} \\
        \hline
        \texttt{madevent}                                                          &  13&G         & 470&G          &  11&T          \\ 
        \texttt{matrix1}                                                           & 3.1&G (23\%)  & 450&G (96\%)   &  11&T (>99\%)  \\ 
        \hspace{1em} \raisebox{1.8pt}{\tikz{\draw[-latex] (0,0.3) |- (0.4,0);}} ext& 450&M (3.4\%) & 3.3&G (<1\%)   & 7.3&G (<1\%)   \\ 
        \hspace{1em} \raisebox{1.8pt}{\tikz{\draw[-latex] (0,0.3) |- (0.4,0);}} int& 1.9&G (14\%)  & 160&G (35\%)   &   2&T (19\%)   \\ 
        \hspace{1em} \raisebox{1.8pt}{\tikz{\draw[-latex] (0,0.3) |- (0.4,0);}} amp& 530&M (4.0\%) & 210&G (44\%)   & 5.5&T (51\%)   \\ 
    \end{tabular}
    \caption{Here we present the number of instructions evaluated by the matrix routine (\texttt{matrix1}) and the total number of instructions evaluated by \texttt{madevent}. In brackets we also present these numbers as a percentage of the \texttt{madevent} total. These results are presented for three processes: $gg\rightarrow t\bar{t}$, $gg\rightarrow t\bar{t}gg$ and $gg\rightarrow t\bar{t}ggg$. We have also broken down the matrix routine into the number of instructions (again alongside their percentage of the total) evaluated by calls to functions that evaluate external spinors (ext), internal wavefunctions (int) and amplitudes (amp). The data in this table was obtained using a combination of Valgrind \cite{valgrind} and KCachegrind \cite{kcachegrind}.}
    \label{tab: instr nohel}
\end{table}

%% file: input/instr_hel.tex

\begin{table*}
    \renewcommand{\arraystretch}{1.5}
    \centering
    \begin{tabular}{ l | r@{}@{}l|c | r@{}@{}l|c | r@{}@{}l|c|}
    & \multicolumn{3}{c|}{\(gg\rightarrow t\bar{t}\)} & \multicolumn{3}{c|}{\(gg\rightarrow t\bar{t}gg\)} & \multicolumn{3}{c|}{\(gg\rightarrow t\bar{t}ggg\)} \\
    \hline
                                                                                   & \multicolumn{2}{c|}{Instructions} & Reduction & \multicolumn{2}{c|}{Instructions} & Reduction & \multicolumn{2}{c|}{Instructions} & Reduction \\\hline
        \texttt{madevent}                                                          &   11&G         & 15\%      & 180&G        & 62\%      &   5&T         & 55\%      \\ 
        \texttt{matrix1}                                                           &    1&G (9.3\%) & 68\%      & 160&G (90\%) & 64\%      & 4.9&T (98\%)  & 55\%      \\ 
        \hspace{1em} \raisebox{1.8pt}{\tikz{\draw[-latex] (0,0.3) |- (0.4,0);}} ext&   76&M (<1\%)  & 83\%      & 100&M (<1\%) & 97\%      & 110&M (<1\%)  & 98\%      \\ 
        \hspace{1em} \raisebox{1.8pt}{\tikz{\draw[-latex] (0,0.3) |- (0.4,0);}} int&  540&M (4.8\%) & 72\%      &  16&G (8.9\%)& 90\%      & 180&G (3.6\%) & 91\%      \\ 
        \hspace{1em} \raisebox{1.8pt}{\tikz{\draw[-latex] (0,0.3) |- (0.4,0);}} amp&  280&M (2.6\%) & 47\%      &  77&G (42\%) & 63\%      & 1.7&T (33\%)  & 69\%     \\ 
    \end{tabular}
    
    \caption{The same as Table (\ref{tab: instr nohel}) but this time with helicity recycling enabled. We also present the percentage reduction between the two tables.}
    \label{tab: instr hel}
\end{table*}

%% file: input/speedup.tex
\begin{table*}
    \renewcommand{\arraystretch}{1.5}
    \begin{subtable}{0.3\textwidth}
        \centering
        \begin{tabular}{ c | c }
            Process & Speed-up \\
            \hline
            \(g g \rightarrow t \bar{t}\) & 1.36\(\times\) \\ 
            \(g g \rightarrow t \bar{t} g\) & 1.43\(\times\) \\ 
            \(g g \rightarrow t \bar{t} g g\) & 2.27\(\times\) \\ 
            \(q q \rightarrow t \bar{t} q q\) & 1.27\(\times\)
        \end{tabular}
        \caption{A selection of $t\bar{t}$ processes.}
        \label{tab: madevent ttbar}
    \end{subtable}\hfill{}
    \begin{subtable}{0.3\textwidth}
        \centering
        \begin{tabular}{ c | c }
            Process & Speed-up \\
            \hline
            \(q q \rightarrow W^+ W^- q q\) & 1.67\(\times\) \\ 
            \(q q \rightarrow W^+ W^- g g\) & 1.89\(\times\) \\ 
            \(g g \rightarrow W^+ W^- q q\) & 1.89\(\times\) \\ 
            \(g q \rightarrow W^+ W^- g q\) & 2.13\(\times\) \\ 
        \end{tabular}
        \caption{ $p p \rightarrow W^+W^-jj$ processes.}
        \label{tab: madevent wpwm}
    \end{subtable}\hfill{}
    \begin{subtable}{0.3\textwidth}
        \centering
        \begin{tabular}{ c | c }
            Process & Speed-up \\
            \hline
            \(q q \rightarrow e^+e^-\) & 1.02\(\times\) \\ 
            \(q q \rightarrow e^+e^-g\) & 1.03\(\times\) \\ 
            \(g g \rightarrow e^+e^-qq\) & 1.09\(\times\) \\ 
        \end{tabular}
        \vspace{.46cm}
        \caption{ A selection of $e^+e^-$ processes.}
        \label{tab: madevent e+e-}
    \end{subtable}
    \caption{The factor speed-up of madevent as a result of using helicity recycling for a selection of SM processes. Here $q \in \{u,d,c,s\}$. One can see two general principles at play. Firstly, the more diagrams a process has and the more complicated those diagrams are, the bigger the speed increase. For example, see the effect of adding more gluons in (a). Secondly, the more helicity combinations we filter away, the lower the speed increase. This is why processes with more fermions see a lower speed-up.}
    \label{tab: madevent}
\end{table*}

%% file: input/tg.tex
\begin{tikzpicture}
	\begin{feynman}
		\vertex (i1);
		\vertex [below right=0.5cm and 2cm of i1] (a);
		\vertex [below =1cm of a] (b);
		\vertex [below =1cm of b] (d);
		\vertex [below =3cm of a] (c);
		\vertex [below left=0.5cm and 2cm of c] (i2);

		\vertex [above right=0.5cm and 2cm of a] (f1);

		\vertex [right=2cm of b] (f3);
		\vertex [right=2cm of d] (f2);
		\vertex [below right=0.5cm and 2cm of c] (f4);

		\vertex [below=1cm of i2] (dummy);

		\diagram* {
		(i1) -- [gluon] (a) ,
		(b) -- [fermion, edge label'=\(t_1\)] (a),
		(b) -- [gluon, edge label=\(t_2\)] (d),
		(c) -- [gluon, edge label'=\(t_3\)] (d),
		(c) -- [gluon] (i2),
		(a) -- [fermion, edge label=\(t\)] (f1),
		(f3) -- [fermion, edge label'=\(\bar{t}\)] (b),
		(f2) -- [gluon] (d),
		(c) -- [gluon] (f4),
		};
	\end{feynman}
\end{tikzpicture}